\documentclass[12pt,english]{article}
\usepackage[T1]{fontenc}
\usepackage[latin1]{inputenc}
\usepackage{babel}
\usepackage{graphics}

\makeatletter

\providecommand{\LyX}{L\kern-.1667em\lower.25em\hbox{Y}\kern-.125emX\@}
\newcommand{\noun}[1]{\textsc{#1}}

\usepackage[T1]{fontenc}
\usepackage{graphicx}
\usepackage{setspace}


\makeatother
\begin{document}
~\\
  \textbf{Title:} Mechanisms of noise-resistance in genetic oscillators\\
 \textbf{Authors:} Jos\'{e} M.~G. Vilar\( ^{1,2} \), Hao Yuan Kueh\( ^{1} \),
Naama Barkai\( ^{3} \), and Stanislas Leibler\( ^{1,2} \)\\
\\
 \( ^{1} \)Howard Hughes Medical Institute, Departments of Molecular
Biology and Physics, Princeton University, Princeton, NJ 08544\\
 \( ^{2} \)The Rockefeller University, 1230 York Avenue, New York,
NY 10021\\
 \( ^{3} \)Departments of Molecular Genetics and Physics of Complex
Systems, Weizmann Institute of Science, Rehovot, Israel\\

\newpage

~\\
 \textbf{Abstract}

A wide range of organisms use circadian clocks to keep internal sense
of daily time and regulate their behavior accordingly. Most of these
clocks use intracellular genetic networks based on positive and negative
regulatory elements. The integration of these {}``circuits'' at
the cellular level imposes strong constraints on their functioning
and design. Here we study a recently proposed model~{[}N.~Barkai
and S.~Leibler, \emph{Nature}, 403:267--268, 2000{]} that incorporates
just the essential elements found experimentally. We show that this
type of oscillator is driven mainly by two elements: the concentration
of a repressor protein and the dynamics of an activator protein forming
an inactive complex with the repressor. Thus the clock does not need
to rely on mRNA dynamics to oscillate, which makes it especially resistant
to fluctuations. Oscillations can be present even when the time average
of the number of mRNA molecules goes below one. Under some conditions,
this oscillator is not only resistant to but paradoxically also enhanced
by the intrinsic biochemical noise.

\newpage

The environment changes in a highly periodic manner. There are, among
others, daily cycles of light and dark as well as annual cycles of
changing climates and physical conditions. Such environmental periodicity
may create the necessity for organisms to develop internal time-keeping
mechanisms to accurately anticipate these external changes and modify
their state accordingly~\cite{edmunds}. In particular, a wide range
of organisms, as diverse as cyanobacteria and mammals, have evolved
circadian rhythms ---biological clocks with a period of about twenty
four hours that evoke and regulate physiological and biochemical changes
to best suit different times of the day.

Recent findings show that the molecular mechanisms upon which these
clocks rely share many common features among species~\cite{dunlap}.
The main characteristic is the presence of intracellular transcription
regulation networks with a set of clock elements that give rise to
stable oscillations in gene expression. A positive element activates
genes coupled to the circadian clock. It simultaneously promotes the
expression of a negative element, which in turn represses the positive
element. The cycle completes itself upon degradation of the negative
element and re-expression of the positive element.

A crucial feature of circadian clocks is the ability to maintain a
constant period over a wide range of internal and external fluctuations~\cite{edmunds}.
Such robustness ensures that the clock runs accurately and triggers
the expression of clock-dependent genes at the appropriate time of
the day. For instance, fluctuations in temperature affect chemical
reaction rates and may perturb oscillatory behavior. Another source
of fluctuations may be the presence of internal noise due to the stochastic
nature of chemical reactions~\cite{barkai}. Low numbers of molecules
may be responsible for random fluctuations that can destabilize the
oscillatory behavior of the biochemical network~\cite{arkin}. Yet,
circadian clocks maintain a fairly constant period amidst such fluctuations.

\noindent ~\\
 \textbf{Description of the model}

\noindent To study possible strategies, or principles, that biological
systems may use to minimize the effect of stochastic noise on circadian
clocks, we examine a minimal model based on the common positive and
negative control elements found experimentally~\cite{barkai}. This
model is described in Figure~\ref{network}. It involves two genes,
an activator \( A \) and a repressor \( R \), which are transcribed
into mRNA and subsequently translated into protein. The activator
\( A \) binds to the \( A \) and \( R \) promoters, which increases
their transcription rate. Thus, \( A \) acts as the positive element
in transcription, whereas \( R \) acts as the negative element by
sequestering the activator.

The deterministic dynamics of the model is given by the set of reaction
rate equations \begin{equation}
\label{rateequations}
\begin{array}{lcl}
{dD_{A}}/{dt} & = & \theta _{A}D_{A}'-\gamma _{A}D_{A}A\\
{dD_{R}}/{dt} & = & \theta _{R}D_{R}'-\gamma _{R}D_{R}A\\
{dD_{A}'}/{dt} & = & \gamma _{A}D_{A}A-\theta _{A}D_{A}'\\
{dD_{R}'}/{dt} & = & \gamma _{R}D_{R}A-\theta _{R}D_{R}'\\
{dM_{A}}/{dt} & = & \alpha _{A}'D_{A}'+\alpha _{A}D_{A}-\delta _{M_{A}}M_{A}\\
{dA}/{dt} & = & \beta _{A}M_{A}+\theta _{A}D_{A}'+\theta _{R}D_{R}'\\
 &  & -A(\gamma _{A}D_{A}+\gamma _{R}D_{R}+\gamma _{C}R+\delta _{A})\\
{dM_{R}}/{dt} & = & \alpha _{R}'D_{R}'+\alpha _{R}D_{R}-\delta _{M_{R}}M_{R}\\
{dR}/{dt} & = & \beta _{R}M_{R}-\gamma _{C}AR+\delta _{A}C-\delta _{R}R\\
{dC}/{dt} & = & \gamma _{C}AR-\delta _{A}C\; ,
\end{array}
\end{equation}
 where the variables and constants are as described in the caption
for Figure~\ref{network}. \textbf{\noun{}}This simple model is
not intended to reproduce the particular details of each organism
but to grasp the properties that the core design confers. As in any
general model, the parameters of the values we use are typical ones.
For instance, the rates for bimolecular reactions are all in the range
of diffusion limited reactions. \textbf{\noun{}}

The preceding equations would be strictly valid in a well-stirred
macroscopic reactor. At the cellular level, a more realistic approach
\noun{}has to consider the intrinsic stochasticity of chemical reactions~\cite{vankampen}.
This can be done by transforming the reaction rates into probability
transition rates and concentrations into number of molecules. One
then obtains the so-called master equation which gives the time evolution
of the probability of having a given number of molecules. There is
no general procedure to solve this type of equation analytically,
but it is the starting point to simulate the stochastic behavior of
the system. The basic idea behind such simulations is to perform a
random walk through the possible states of the system, which are defined
by the numbers of molecules of the different reacting species. Starting
from a state with given numbers of molecules, the probability of jumping
to other state with different numbers of molecules (i.e. the probability
for an elementary reaction to happen) can be computed from the master
equation. One can pick up a state and the jumping time according to
that probability distribution, then consider the resulting state as
a new initial state, and repeat this procedure until some final state
or time is reached. In this way, the numbers of molecules change in
time with the statistical properties given by the master equation.
There are several algorithms to implement this. The main difference
among them is the specific way in which they compute the probabilities
and select the states. For chemical reactions with few components,
it is customary to use the so-called \noun{}Gillespie algorithm~\cite{gillespie}. \textbf{\noun{}}

This intrinsic probabilistic behavior in the evolution of the number
of molecules gives rise to fluctuations that are usually referred
to as noise. In general, the term noise is used for any disturbance
interfering with a signal or with the operation of system. In the
case of chemical reactions, the signal would be the average production
of the reacting species whereas the disturbance would arise as a consequence
of the fluctuations around that average value. We use term noise rather
than fluctuations to emphasize the disturbing effect that these fluctuations
can have. Thus, although related, both terms do not mean the same.
For instance, there can be large fluctuations in some molecular species
but, if their characteristic time is very short compared to those
of other processes that take place, they would introduce little noise. 

In Figure~\ref{oscillations} we compare the results of the stochastic
and deterministic approaches. We show the levels of \( A \) protein
and \( R \) protein over time for the set of parameter values and
initial conditions given in the caption of Figure~\ref{network}.
The deterministic results were obtained from numerical integration
of Eqs.~(\ref{rateequations}), whereas the stochastic results were
obtained by computer simulation using the Gillespie algorithm. The
main difference between the deterministic and stochastic time courses
is the presence of random fluctuations in the latter. In the deterministic
model every circadian cycle is identical to the previous one. The
stochastic model shows some variability in the numbers of molecules
and the period length, corresponding to the intrinsic fluctuations
of the biochemical network. For these values of the parameters, both
stochastic and deterministic approaches give similar results. We have
also used different initial conditions and in all the cases we have
observed that the behavior of the long term solution is the same.

\noindent ~\\
 \textbf{Model simplification}

\noindent To gain further insight into the essential elements that
are responsible for the oscillations, we will simplify as much as
possible the deterministic rate equations. By making various quasi-steady
state assumptions~\cite{murray}, it is possible to simplify the
set of Eqs.~(\ref{rateequations}) into a two variable system with
the repressor \( R \) and the complex \( C \) as the two slow variables:
\textbf{\begin{eqnarray}
\frac{dR}{dt} & = & \frac{\beta _{R}}{\delta _{M_{R}}}\frac{\alpha _{R}\theta _{R}+\alpha _{R}'\gamma _{R}\widetilde{A}(R)}{\theta _{R}+\gamma _{R}\widetilde{A}(R)}-\gamma _{C}\widetilde{A}(R)R+\delta _{A}C-\delta _{R}R\nonumber \\
\frac{dC}{dt} & = & \gamma _{C}\widetilde{A}(R)R-\delta _{A}C\label{slow} 
\end{eqnarray}
}

\noindent where\[
\widetilde{A}(R)={1\over2 }({\alpha _{A}'\rho (R)}-K_{d})+{1\over2 }\sqrt{({\alpha _{A}'\rho (R)}-K_{d})^{2}+4{\alpha _{A}\rho (R)}K_{d}}\]
 with \( \rho (R)=\beta _{A}/\delta _{M_{A}}(\gamma _{C}R+\delta _{A}) \)
and \( K_{d}=\theta _{A}/\gamma _{A} \). Notice that the nonlinearity
in the equations enters trough the quasi-equilibrium value of \( A \),
\( \widetilde{A}(R) \), which is a function of \( R \). The main
idea behind these approximations is that there are fast and slow variables.
Fast variables are assumed to be at an effective equilibrium whereas
slow variables are responsible for the dynamics of the system. Thus,
given the set of Eqs.~(\ref{rateequations}), we assume that all
the derivatives except \( dR/dt \) and \( dC/dt \) are zero. 

In Figure~\ref{versus} we show that for the values of the parameters
we use the numerical solutions for the trajectories of the two-variable
system {[}Eqs.~(\ref{slow}){]} agree closely with the solutions
of the full system {[}Eqs.~(\ref{rateequations}){]}, except for
quantitative differences in the peak levels and times at the beginning
of each cycle. These differences arise because the time scale separation
between fast and slow variables is not sufficiently large for quasi-steady
state assumptions to be exact. These results indicate, nevertheless,
that the dynamics of the system is mainly determined by two component
concentrations: those of the complex and the repressor. The other
components are driven mainly by these two elements and their effects
enter the system through effective parameters. It is worth emphasizing
that the reduced two-variable model is aimed just to offer insights
into the qualitative behavior of the system and to show how the properties
that one observes in the full system are already present in a simple
two variable model. Thus, whenever we present simulation results for
the deterministic system, except if otherwise stated, we are referring
to the full system. Regarding the validity of the two-variable model,
it is a good approximation when the dynamics of mRNA and the activator
is faster than that of the complex and repressor. For instance, it
will remain valid if \( \delta _{R} \) is decreased or some of \( \delta _{A} \),
\( \delta _{M_{A}} \), \( \delta _{M_{R}} \), \( \theta _{A} \),
and \( \theta _{R} \) are increased with respect to the parameters
of the caption for Figure~\ref{network}. 

\noindent ~\\
 \textbf{Limit cycle oscillations and stability analysis}

\noindent The existence of oscillations in the two-variable system
can be inferred from application of the Poincar\'{e}-Bendixson theorem~\cite{strogatz}.
This theorem states that a two-dimensional system of the type we are
considering exhibits limit cycles if it is confined in a closed bounded
region that does not contain any stable fixed points. The trajectory
of the system is confined since the number of molecules cannot reach
infinite values. The fixed points and their stability can be determined
by following a standard linear stability analysis. There is a single
fixed point for positive concentrations. In our case, the signs of
the real parts of the eigenvalues of the matrix describing the linearized
dynamics around this point are given by \begin{equation}
\label{tracetext}
\tau =\left| \frac{\partial \widetilde{A}(R)}{\partial R}\right| \cdot \left[ \gamma _{C}R-\frac{\beta _{R}}{\delta _{M_{R}}}\frac{(\alpha _{R}'-\alpha _{R})\theta _{R}\gamma _{R}}{[\theta _{R}+\gamma _{R}\widetilde{A}(R)]^{2}}\right] -[\gamma _{C}\widetilde{A}(R)+\delta _{R}+\delta _{A}]\; ,
\end{equation}
 where \( \tau  \) is the trace of the matrix. All the quantities
are evaluated at the fixed point \( (R_{0},C_{0}) \). When \( \tau  \)
is positive, the real part of the eigenvalues is positive, the fixed
point is unstable, and there is a limit cycle in the system, which
gives sufficient conditions for the existence of oscillations. Evaluation
of Eq.~(\ref{tracetext}) shows that \( \tau  \) is indeed positive
for the set of parameters we are using. The domain in which \( \tau  \)
is positive is rather broad. For instance, the function \( \tau  \)
remains positive when \( \theta _{A} \) and \( \theta _{R} \) are
multiplied by a factor \( K \) with \( 0.024<K<10.7 \); when all
transcription rates (\( \alpha  \) and \( \alpha ' \)) are multiplied
by \( K \) with \( K>0.08 \); or when protein and mRNA degradation
rates are multiplied by \( K \) with \( 0.0009<K<3.5 \). \textbf{\noun{}}When
\( \tau  \) is negative, the fixed point is stable and, in principle,
no conclusion about the existence of limit cycles can be drawn. \textbf{\noun{}}In
this case, the presence of oscillations could also depend on the initial
conditions. For the full model, the ranges of parameters that give
rise to oscillations are not exactly the same but remain very close
to the previous ones.

The mechanism responsible for oscillations is illustrated in Figure~\ref{phaseplane}
through the phase portrait of the two-variable model. Starting with
low numbers of initial molecules near the origin of the phase plane,
the trajectory rapidly shoots upwards along the \( \dot{R}=0 \) nullcline
(the dot over a variable means its time derivative). Here, the high
levels of \( A \), present due to auto-activated transcription, rapidly
induce the formation of the complex \( C \). Reaching the maximum
of the nullcline, the trajectory `falls off' the edge and moves rapidly
diagonally right and downwards, corresponding to a drop in \( C \)
and a rise in \( R \). The trajectory curves around the \( \dot{R}=0 \)
nullcline and hits the \( \dot{C}=0 \) nullcline, where it slowly
returns to the left and approaches the fixed point \( (R_{0},C_{0}) \).
When approaching the fixed point, \( \dot{R} \) decreases sharply,
taking the trajectory past the fixed point and sending it back upwards
to initiate a new cycle.

The trajectory in Figure~\ref{phaseplane} comprises a fast phase
corresponding to the rapid production of \( C \) and \( R \), and
a slow phase corresponding to the slow degradation of \( R \). These
two distinct phases are characteristic of excitable systems, the classic
example of which is the Fitz Hugh-Nagumo model for action potential
transmission in neurons~\cite{kaplan,keener}. The fast and slow
leg correspond to the excitable and refractory phase of the system,
respectively. Thus, the system oscillates as it avoids the fixed point
\( (R_{0},C_{0}) \) and hits the \( \dot{R}=0 \) nullcline on the
left to begin the excitable phase of a new cycle.

As we have already pointed out, the deterministic analysis can be
useful to grasp the main properties of the system under certain conditions.
Unfortunately such conditions are not known \textit{a priori} without
a stochastic analysis. Surprisingly enough, we have found that parameter
values that give rise to a stable steady state in the deterministic
limit continue to produce reliable oscillations in the stochastic
case, as shown in Figure~\ref{josestheman}. Therefore, the presence
of noise not only changes the behavior of the system by adding more
disorder, but can also lead to marked qualitative differences.

How can the system continue to produce oscillations even when deterministic
rate equations predict a stable steady state? The system always evolves
towards a stable fixed point, as sketched in Figure~\ref{oneosc}.
However, a perturbation of sufficient magnitude near the fixed point,
e.g. as illustrated by the dotted arrow in Figure~\ref{oneosc} (notice
that the figure has not been drawn to scale and that the size of the
arrow is not representative of the actual size of the perturbation\noun{)},
may initiate a new cycle. For low numbers of molecules, the intrinsic
fluctuations of chemical reactions can be large enough to continually
send the system into the fast phase after each cycle and thus produce
sustained oscillations. In the deterministic limit (or close to it),
there are no perturbations (or the perturbations are too small) to
initiate a new cycle and the trajectory stays close to the fixed point.
In this case, the system performs better if enough noise is present
in the system. This situation is analogous to that observed in the
Fitz Hugh--Nagumo model, where an optimal amount of noise maximizes
the reliability of the oscillations~\cite{pikovsky}. It is important
to realize that the effects that noise may have on non-linear systems
can be difficult to predict and rather paradoxical~\cite{nsn}. Therefore,
the smaller number of molecules does not necessarily imply more irregular
behavior of the system, as one might intuitively assume~\cite{arkin,johan}.

\noindent ~\\
 \textbf{Significant parameters and noise resistance}

\noindent The mechanism responsible for oscillations involves only
two variables. This has some important consequences for the functioning
of the clock. If we consider the deterministic limit, a two-dimensional
dynamical system of this type either oscillates regularly or does
not oscillate at all. In two dimensions, since trajectories cannot
cross, fixed points and periodic orbits are the only possible attractors.
Other more complicated behaviors such as chaos or quasi periodicity
are not allowed~\cite{strogatz}. \textbf{\noun{}}On the other
hand, the intrinsic stochastic fluctuations of the remaining variables
are effectively averaged and do not significantly affect the performance
of the system.

For instance, one variable that usually plays a prominent role in
many circadian rhythm models is the number of mRNA molecules~\cite{goldbeter}.
In our case, however, mRNA does not enter directly into the dynamics.
It is just an intermediate step in the production of the proteins.
Thus, if \textbf{\noun{}}protein production remains unaltered the
system will oscillate regardless of the number of mRNA molecules involved.
This can be achieved, for instance, by increasing simultaneously the
degradation rates of mRNA and the translation rate of the proteins.
In the deterministic limit of the two-variable model this has no effect
on protein dynamics at all. In the stochastic case the effects are
negligible. Figure~\ref{megacorre} shows the time evolution of repressor
mRNA and protein levels in the system for \( \beta _{A} \), \( \beta _{R} \),
\( \delta _{M_{A}} \) and \( \delta _{M_{R}} \) multiplied by a
factor of hundred. The system essentially alternates between having
zero and one mRNA molecule in the cycle, and the proteins continue
to exhibit remarkably good oscillations.

There are also \textbf{\noun{}}parameters that do affect the properties
of the oscillations. In the deterministic limit, oscillations are
always regular (provided that the two-variable model is a good approximation).
When fluctuations are taken into account, the reliability of the oscillations
may depend on those parameters. One such parameter, as we have seen
in the previous section, is the repressor degradation rate \( \delta _{R} \).
This parameter affects the period of the oscillations (compare e.g.
Figures~\ref{oscillations} and \ref{josestheman}) and also the
stability of the fixed point. One can infer from Eq.~(\ref{tracetext})
that for high or low values of \( \delta _{R} \) the fixed point
becomes stable. In such cases, the deterministic system stops oscillating
but this does not need to be so for the stochastic one, which may
continue producing reliable oscillations.

Notice that the positive feedback is a key element in the clock dynamics.
Its most obvious use is the generation of the instability that gives
rise to oscillations. But it has other not so obvious role that is
closely related to the resistance to noise. In general, gene regulation
is a slow process (with typical characteristic times of about one
hour) and, as a such, is prone to be affected by fluctuations. This
problem gets even worse if the dynamics relies in several coupled
transcriptional feedbacks since the effects of the fluctuations are
then amplified. The positive feedback gives a fast transcriptional
switch, allowing to move fast from low to high transcription rates.
In this way, the time in which the system is prone to fluctuations
is greatly reduced. \textbf{\noun{}}

\noindent ~\\
 \textbf{Conclusions}

\noindent The presence of noise in transcriptional and enzymatic networks
is a fundamental consequence of the stochastic nature of biochemical
reactions. The ability to function effectively and consistently amidst
such random fluctuations is a major issue in gene expression and network
behavior. In this paper we have studied how different factors affect
a simple model for circadian rhythms that exhibits noise resistance.
We found that the oscillations in this model are driven mainly by
two components: a repressor protein and an activator-repressor complex.
This fact is responsible for the reliability of the oscillations.
First, a two-dimensional dynamical system of this kind has a very
simple behavior: in the deterministic limit, it either oscillates
or goes to a steady state. Second, noise and perturbations in the
other variables affect the system only slightly since they do not
enter directly the dynamics. Finally, resistance to noise is achieved
as the number of molecules of any of the two key components reaches
small values only for short periods of time or when they are not driving
the dynamics of the system. In this way, even though some molecular
species may be present in very low numbers, the intrinsic stochasticity
of biochemical reactions can be bypassed.

It is important to emphasize that organisms have evolved networks
to function in the extremely noisy cellular environment. Suitable
network designs, as those that are now emerging from the experimental
data~\cite{barkai,smolen}, can confer resistance against such noise.
In addition, some of these networks may not only be resistant to but
could also be taking advantage of the cellular noise to perform their
functions under conditions in which it would not be possible by deterministic
means.

\noindent ~\\
 \textbf{Acknowledgements}

\noindent This work was supported by the U.S. National Institutes
of Health.

\newpage

\noindent \textbf{\large Figure Captions}{\large \par}

\begin{list}{FIGURE \arabic{figure}:}{\usecounter{figure}
\leftmargin 0 cm
\rightmargin 0 cm
\itemindent 1.5cm
}
\item \label{network} Biochemical network of the circadian oscillator model.
\(D_{A}'\) and
\(D_{A}\) denote the number of activator genes with and without
\(A\) bound to its
promoter respectively;
similarly, \(D_{R}'\) and \(D_{R}\) refer to the repressor
promoter; \(M_{A}\) and \(M_{R}\) denote mRNA of \(A\) and \(R\);
\(A\) and \(R\) correspond to the activator and repressor proteins; and \(C\) to the
the inactivated complex formed by \(A\) and \(R\). The constants \(\alpha\) and
\(\alpha'\) denote the
basal and activated rates of transcription, \(\beta\) 
the rates of translation,  \(\delta\) the rates of spontaneous
degradation, \(\gamma\) the rates of binding of \(A\) to
other components, and \(\theta\) the rates of unbinding of
\(A\) from those components.
Except if otherwise stated, along this
paper we have assumed the following values for the reaction rates: 
\(\alpha_{A}=50 \textrm{ hr}^{-1}\), \(\alpha_{A}' = 500
\textrm{ hr}^{-1}\), \(\alpha_{R}=0.01 \textrm{ hr}^{-1}\), \(\alpha_{R}' = 50
\textrm{ hr}^{-1}\), \(\beta_{A}=50 \textrm{ hr}^{-1}\), \(\beta_{R}= 5
\textrm{ hr}^{-1}\), \(\delta_{MA}=10 \textrm{ hr}^{-1}\),
\(\delta_{MR}=0.5 \textrm{ hr}^{-1}\), \(\delta_{A}=1 \textrm{ hr}^{-1}\),
\(\delta_{R}=0.2 \textrm{ hr}^{-1}\), \(\gamma_{A}=1 \textrm{ mol}^{-1}\textrm{
hr}^{-1}\),
\(\gamma_{R}=1 \textrm{ mol}^{-1}\textrm{ hr}^{-1}\), \(\gamma_{C} = 2 \textrm{
mol}^{-1}
\textrm{ hr}^{-1}\), \(\theta_{A}=50 \textrm{ hr}^{-1}\), \(\theta_{R} = 100
\textrm{ hr}^{-1}\), where \(\textrm{mol}\) means number of molecules.
The initial conditions are \(D_{A}=D_{R}=1 \textrm{ mol}\),
\(D_{A}'=D_{R}'=M_{A}=M_{R}=A=R=C=0\),
which require that the cell has a single copy of the activator and
repressor genes: \(D_{A}+D_{A}'=1 \textrm{ mol}\) and \(D_{R}+D_{R}'=1 \textrm{ mol}\).
The cellular volume is assumed to be the unity so that concentrations and
number of molecules are equivalent.
Notice that we assume that the complex breaks into \(R\) due to the degradation of \(A\) and therefore the parameter \(\delta_{A}\) appears twice in the model. 

\item \label{oscillations} Oscillations in repressor and activator protein
numbers obtained from numerical
simulations of the deterministic (\(a\),\(b\))  and stochastic (\(c\),\(d\))
descriptions of the model. 

\item \label{versus} Time evolution of the quantities \(R\) (continuous line
line) and \(C\) (dashed line) for the system reduced to two variables (\(a\)) by
various quasi-steady state assumptions and for the complete system (\(b\)).

\item \label{phaseplane}
Phase portrait of the two variable oscillator [Eqs.~(\ref{slow})] for the
parameter
values given in the caption for Figure~\ref{network} (the drawing is not to scale).
The thick line illustrates the trajectory of system. \((R_{0}, C_{0})\) is the fixed
point
of the system, and \(\dot{R}\equiv{dR / dt}=0\) and \(\dot{C}\equiv{dC / dt}=0\)
are the \(R\) and \(C\) nullclines
respectively. The solid arrows give the orientation of the direction field on
the nullclines. 

\item \label{josestheman} Time evolution of \(R\) for the deterministic [Eqs.~(\ref{rateequations})] (\(a\))
and stochastic (\(b\))
versions of the model. The values of the parameters are as in
the caption of Figure~\ref{network}
except that now we set \(\delta_{R}=0.05 \textrm{ hr}^{-1}\). For these parameter
values \(\tau < 0\)
so that the fixed point is stable.

\item \label{oneosc}
Phase portrait as in Figure~\ref{phaseplane} but for a situation in which the
system falls into the
stable fixed point \((R_{0},C_{0})\). The dotted arrow to the left of the fixed
point illustrates a perturbation that would initiate a single sweep of the (former) oscillatory trajectory. 

\item \label{megacorre} Stochastic time evolution
of the number of activator (\(a\)) and repressor (\(c\)) molecules; and
activator (\(b\)) and repressor (\(d\)) mRNA molecules. 
 The values of the parameters are as in the caption for Figure~\ref{network}
but now with \(\beta_{A}=5000 \textrm{ hr}^{-1}\), \(\beta_{R}= 500
\textrm{ hr}^{-1}\), \(\delta_{MA}=1000 \textrm{ hr}^{-1}\), and
\(\delta_{MR}=50 \textrm{ hr}^{-1}\).

\end{list}
\newpage \pagestyle{empty} ~ \vfill
~ \centering
\resizebox*{8cm}{!}{\includegraphics{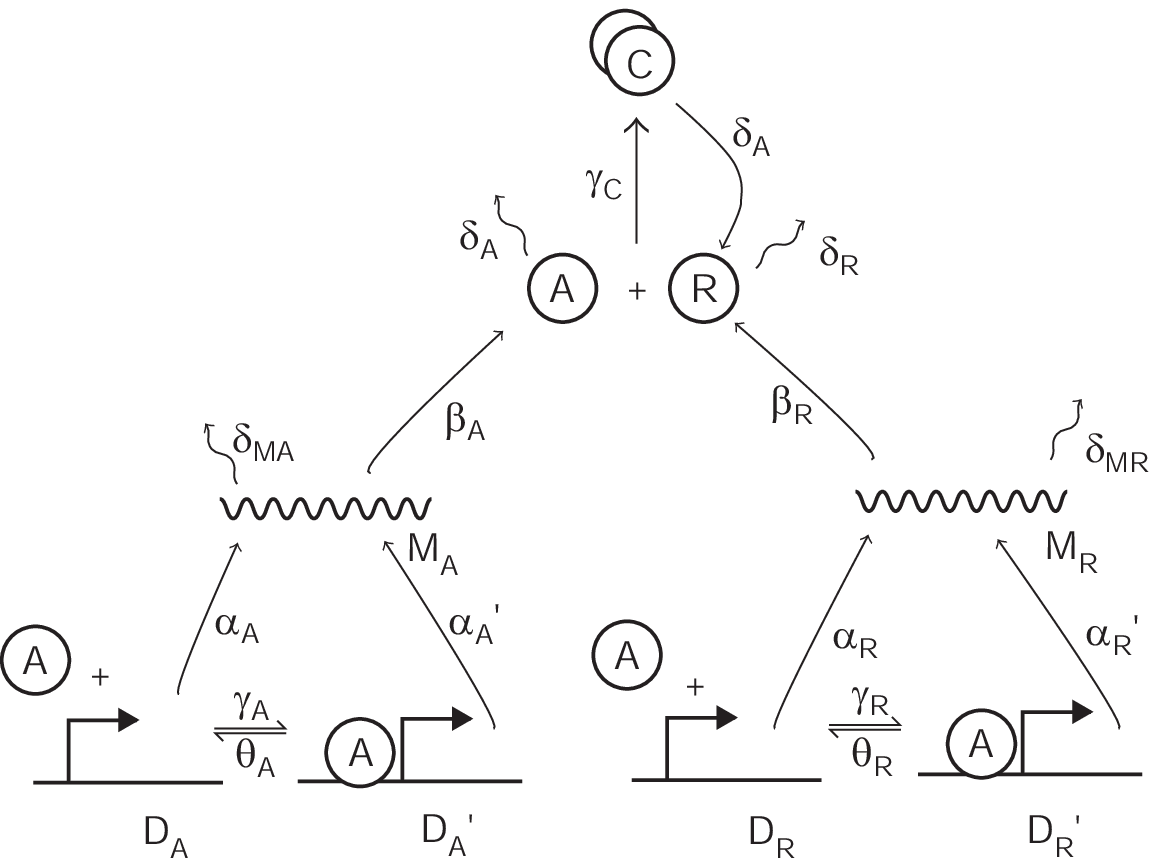}}  ~ \vfill
~ \textbf{FIGURE 1}

\newpage \pagestyle{empty} ~ \vfill
~ \centering
\resizebox*{7.5cm}{!}{\includegraphics{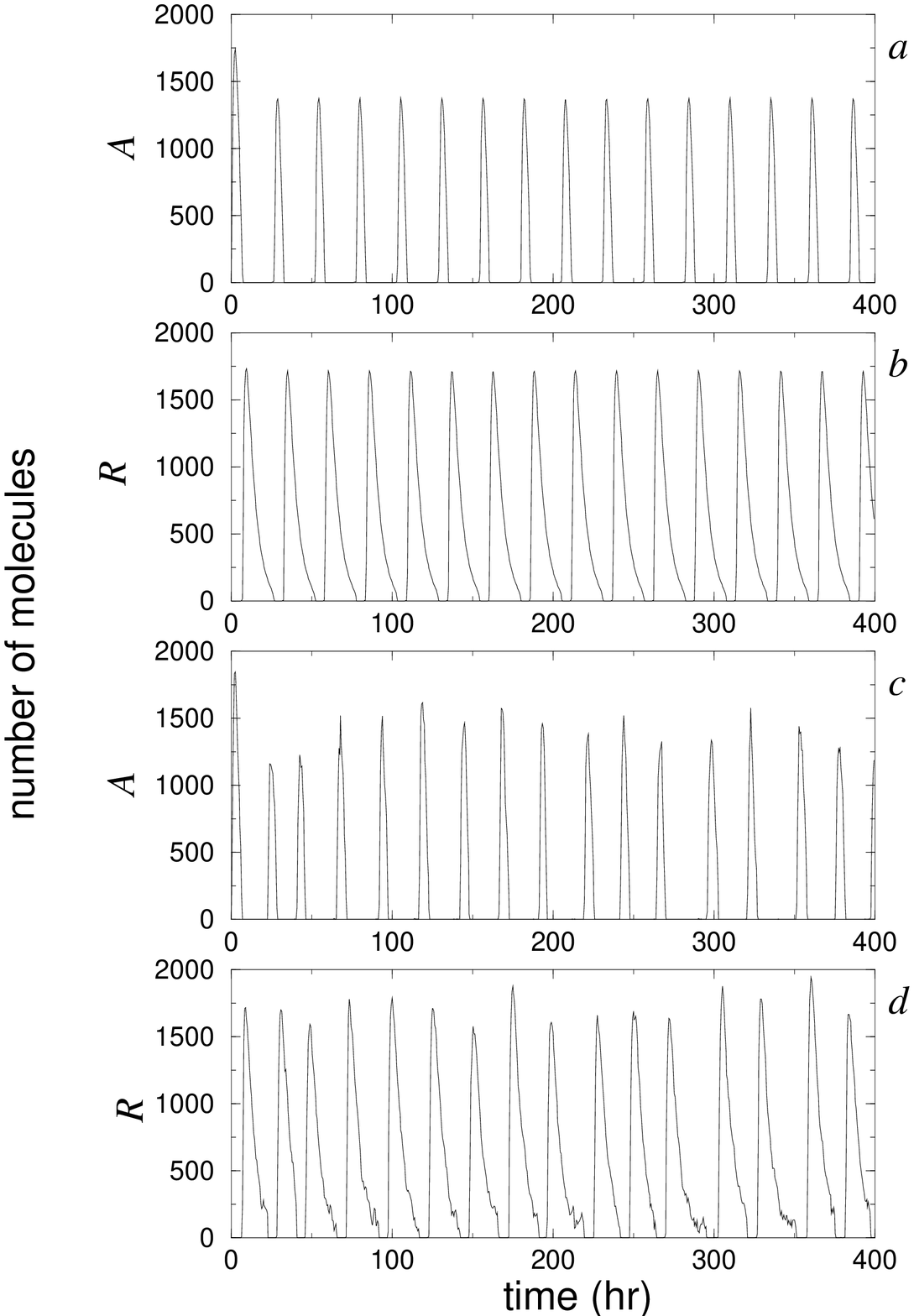}}  ~ \vfill
~ \textbf{FIGURE 2}

\newpage \pagestyle{empty} ~ \vfill
~ \centering
\includegraphics[width=7.5 cm]{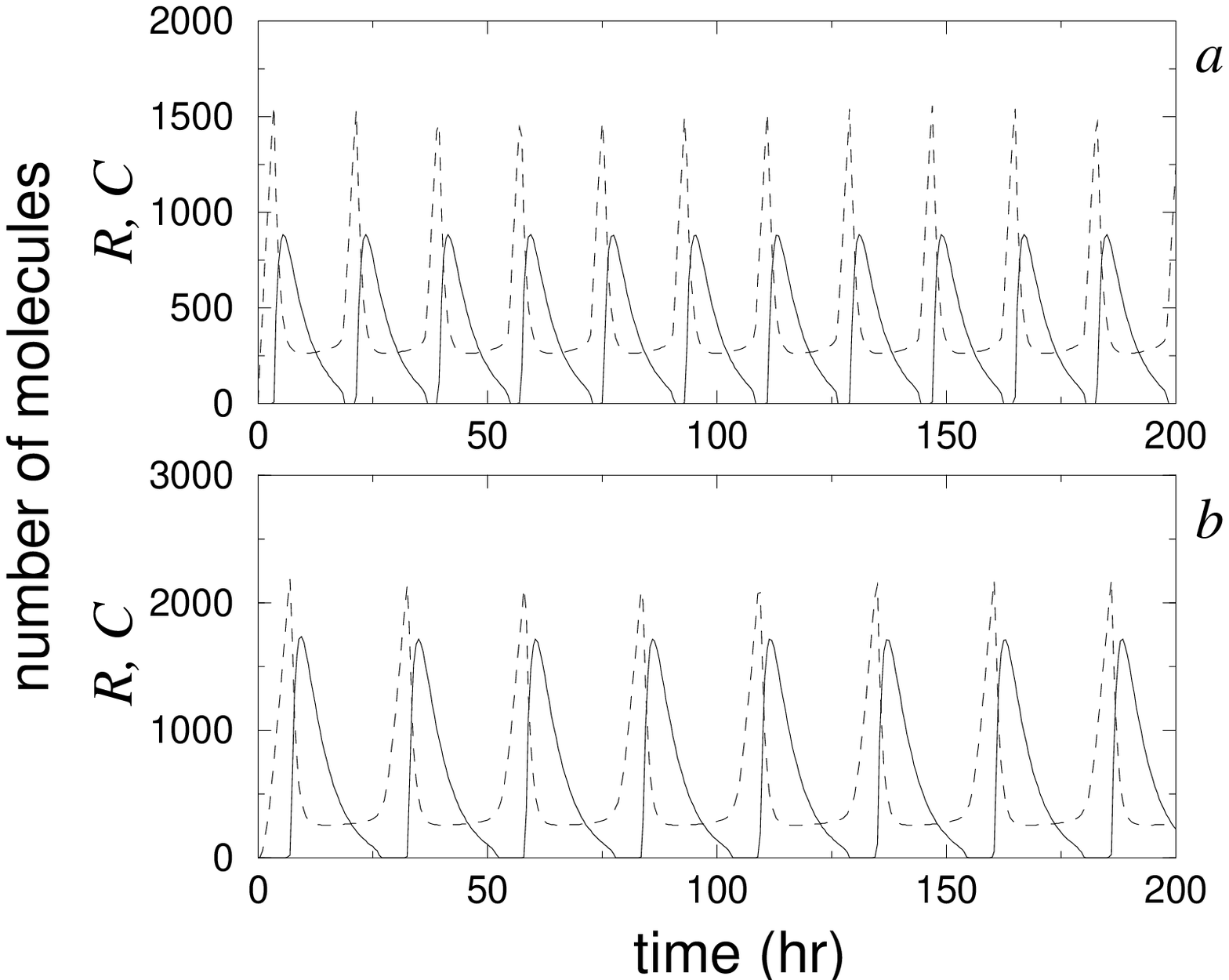} ~ \vfill
~ \textbf{FIGURE 3}

\newpage \pagestyle{empty} ~ \vfill
~ \centering
\includegraphics[width=7 cm]{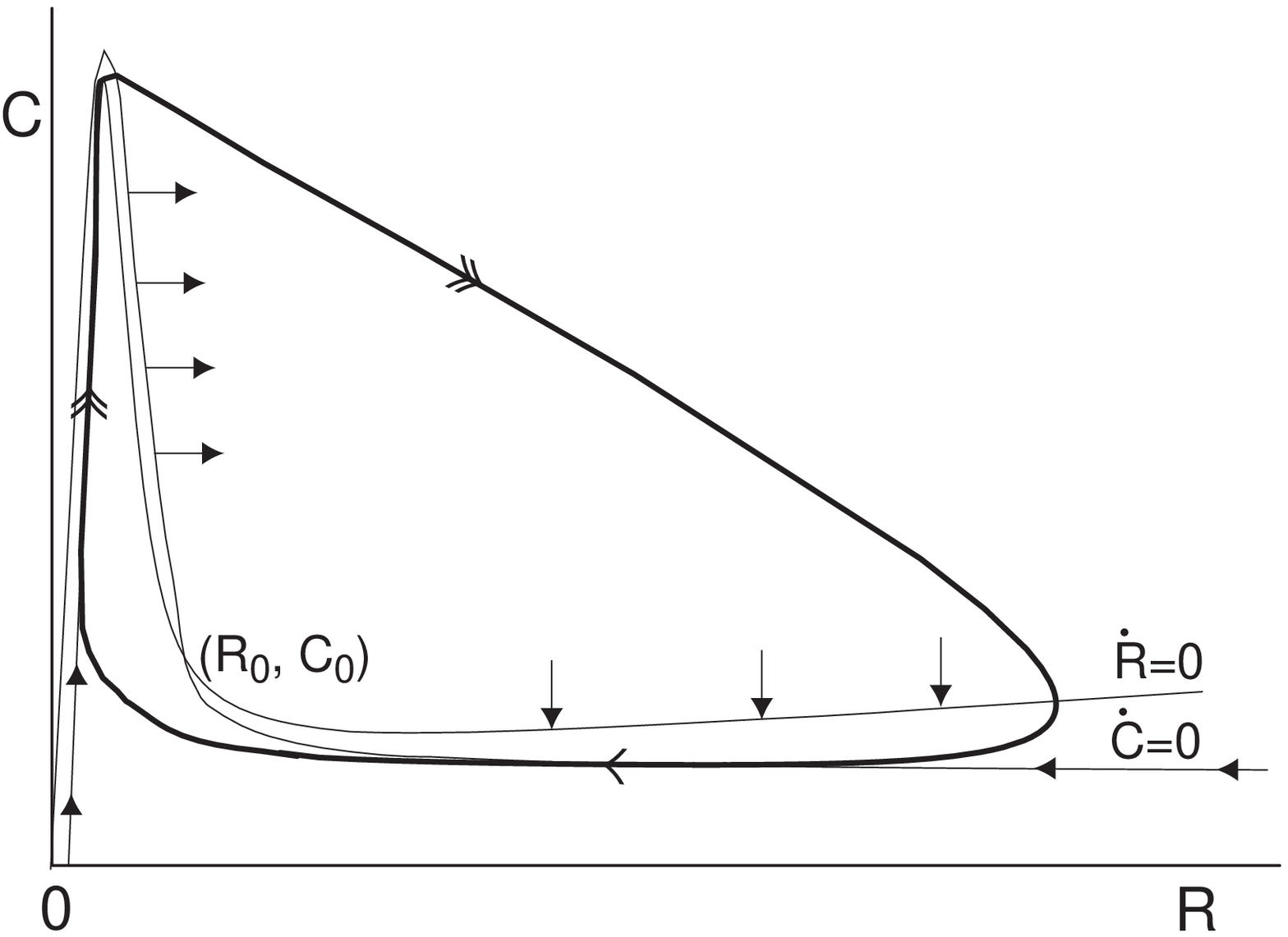} ~ \vfill
~ \textbf{FIGURE 4}

\newpage \pagestyle{empty} ~ \vfill
~ \centering
\resizebox*{7.5cm}{!}{\includegraphics{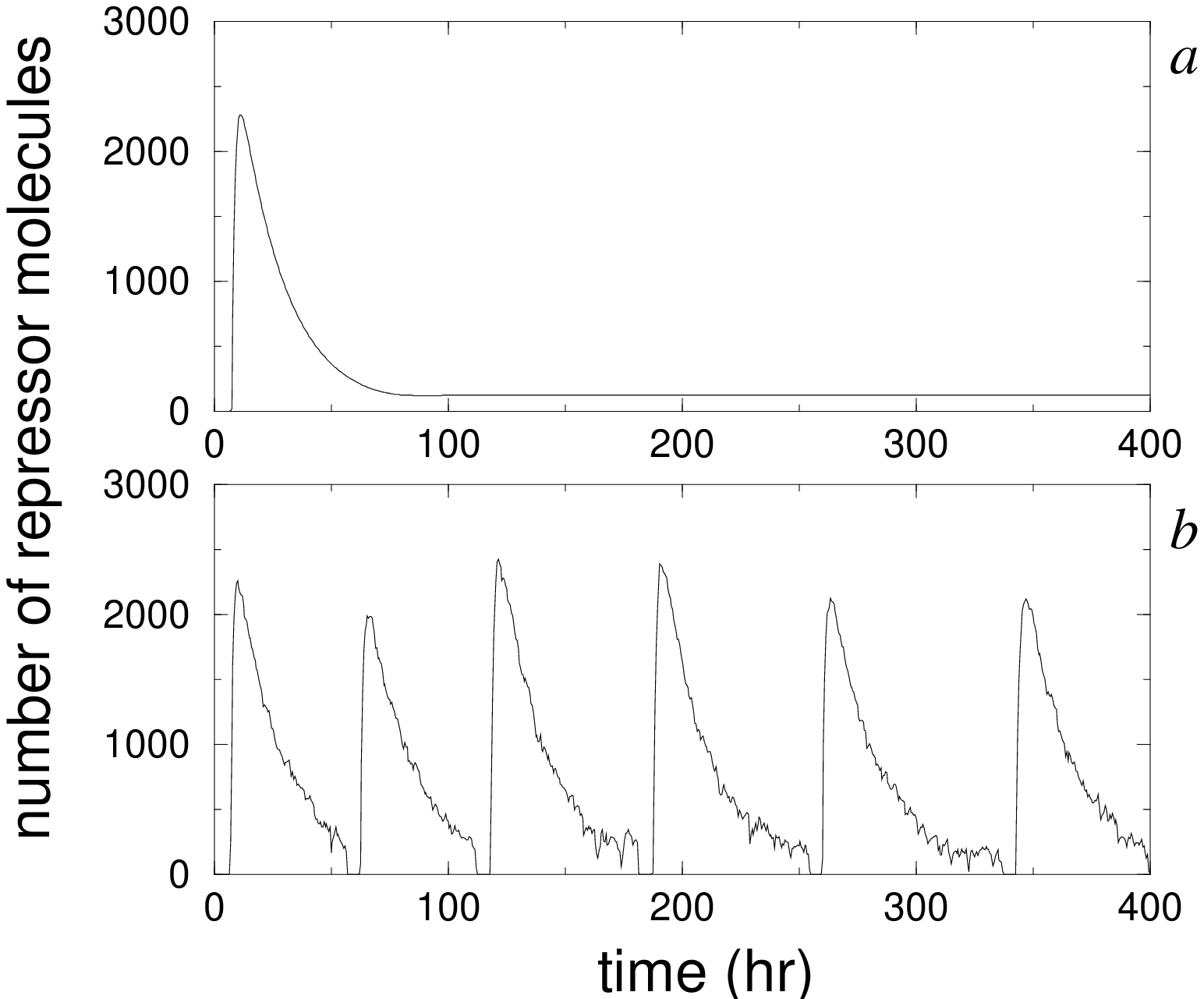}}  ~ \vfill
~ \textbf{FIGURE 5}

\newpage \pagestyle{empty} ~ \vfill
~ \centering
\includegraphics[width=7 cm]{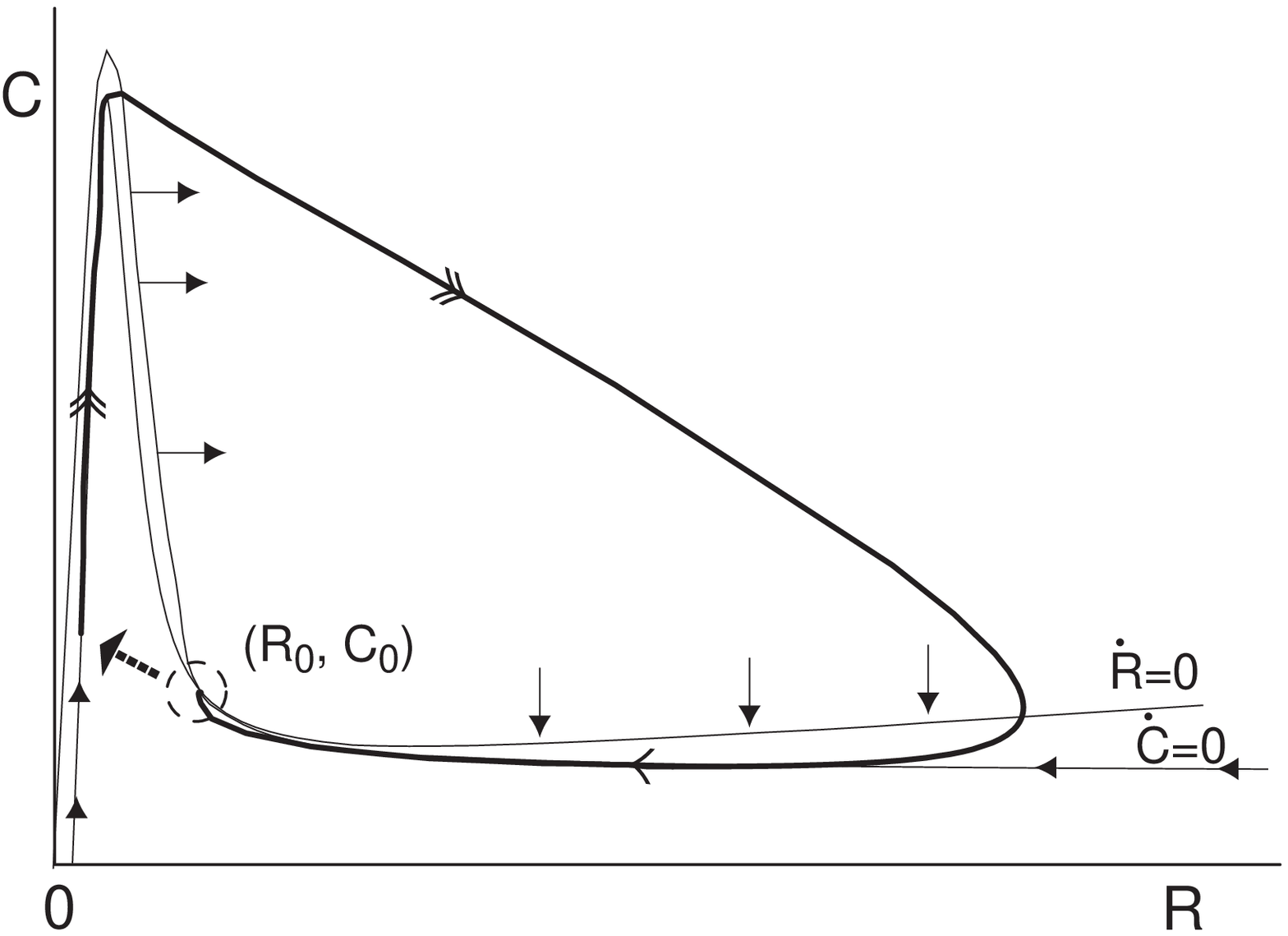} ~ \vfill
~ \textbf{FIGURE 6}

\newpage \pagestyle{empty} ~ \vfill
~ \centering
\includegraphics[width=7.5 cm]{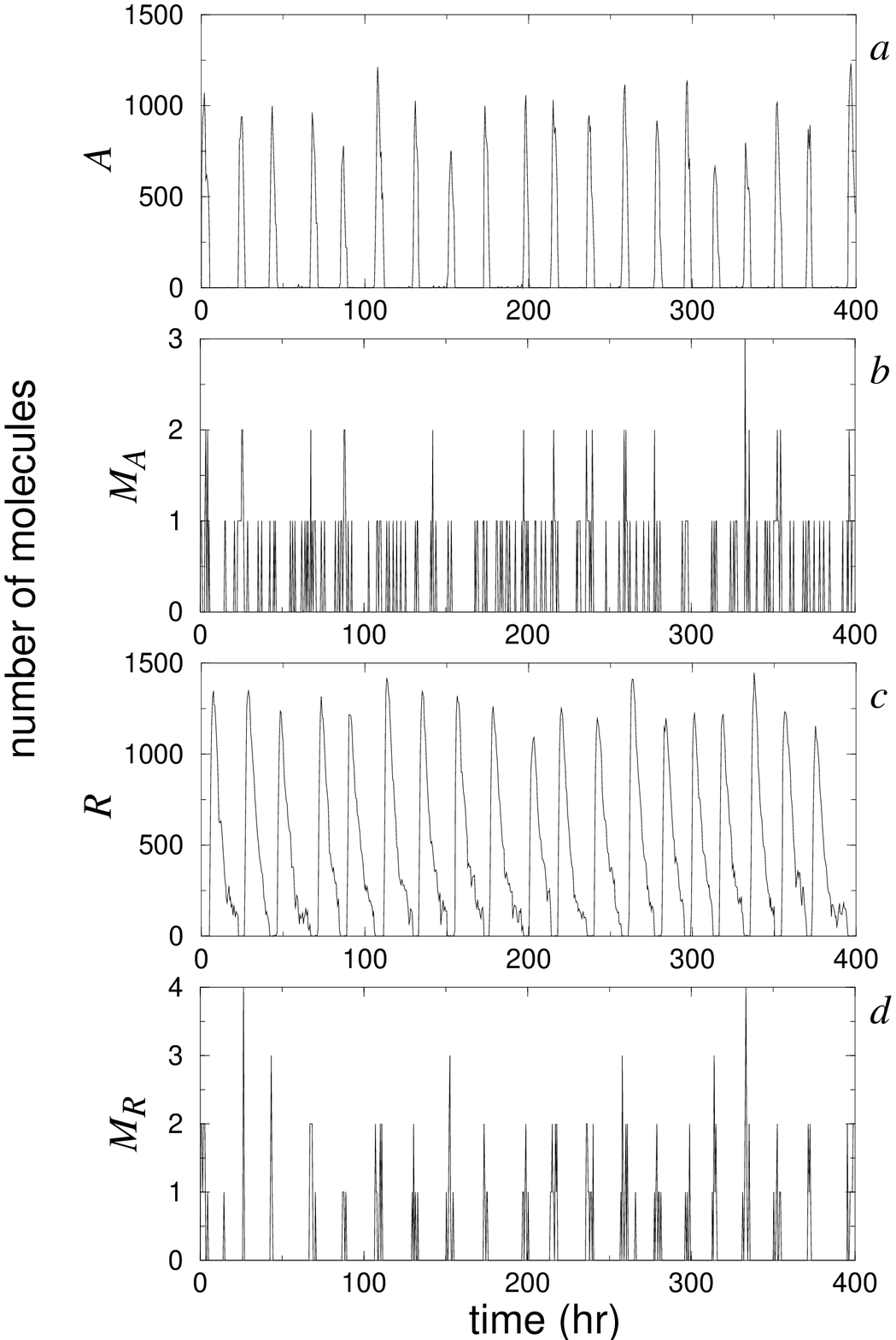} ~ \vfill
~ \textbf{FIGURE 7}
\end{document}